\documentclass[conference]{IEEEtran}
\IEEEoverridecommandlockouts
\usepackage{cite}

\usepackage{amsmath,amssymb,amsfonts}
\usepackage{hyperref}
\hypersetup{colorlinks=true, unicode=true, linkcolor=[rgb]{0.10,0.05,0.67}, citecolor=[rgb]{0.10,0.05,0.67}, filecolor=[rgb]{0.10,0.05,0.67}, urlcolor=[rgb]{0.10,0.05,0.67}}
\usepackage{algorithm}
\usepackage{algpseudocode}
\usepackage{graphicx}
\usepackage{textcomp}
\usepackage{xcolor}
\def\BibTeX{{\rm B\kern-.05em{\sc i\kern-.025em b}\kern-.08em
    T\kern-.1667em\lower.7ex\hbox{E}\kern-.125emX}}

\begin{document}
\title{CRYPTO-MINE: Cryptanalysis via \\Mutual Information Neural Estimation\vspace{-2mm}}


\author{Benjamin D. Kim, Vipindev Adat Vasudevan, Jongchan Woo, Alejandro Cohen, \\ Rafael G. L. D'Oliveira, Thomas Stahlbuhk, and Muriel M\'edard
\thanks{\copyright~2023 IEEE. Personal use of this material is permitted. Permission from IEEE must be obtained for all other uses, in any current or future media, including reprinting/republishing this material for advertising or promotional purposes, creating new collective works, for resale or redistribution to servers or lists, or reuse of any copyrighted component of this work in other works.}
\thanks{This work was supported by MIT Lincoln Laboratory.} 
\thanks{Benjamin D. Kim and Vipindev Adat Vasudevan contributed equally to this work. B.~D.~Kim is with University of Illinois Urbana-Champaign, USA (e-mail: bdkim4@illinois.edu). J.~Woo, V.~Adat Vasudevan and M.~Médard are with Massachusetts Institute of Technology, USA (email: \{jc\_woo, vipindev, medard\}@mit.edu). A.~Cohen is with Technion, Israel (e-mail: alecohen@technion.ac.il). R.~G. L. D'Oliveira is with Clemson University, USA (e-mail: rdolive@clemson.edu). T.~Stahlbuhk is with MIT Lincoln Laboratory, USA (e-mail: thomas.stahlbuhk@ll.mit.edu).}}

\maketitle
\begin{abstract}
The use of Mutual Information (MI) as a measure to evaluate the efficiency of cryptosystems has an extensive history. However, estimating MI between unknown random variables in a high-dimensional space is challenging. Recent advances in machine learning have enabled progress in estimating MI using neural networks. This work presents a novel application of MI estimation in the field of cryptography. We propose applying this methodology directly to estimate the MI between plaintext and ciphertext in a chosen plaintext attack. The leaked information, if any, from the encryption could potentially be exploited by adversaries to compromise the computational security of the cryptosystem.  We evaluate the efficiency of our approach by empirically analyzing multiple encryption schemes and baseline approaches. Furthermore, we extend the analysis to novel network coding-based cryptosystems that provide individual secrecy and study the relationship between information leakage and input distribution.
\end{abstract}


\section{Introduction}



Mutual information (MI) has a long history in cryptography, dating back to the era of Claude Shannon, who introduced the concept of information leakage and its relationship to secure communication systems \cite{shannon1949communication}. According to the definition of perfect secrecy \cite{shannon1949communication}, the MI between the ciphertext and the plaintext is zero (i.e., \emph{a posteriori} probability and \emph{a priori} probability of finding the plaintext remains the same even if the ciphertext is known). However, achieving perfect secrecy requires as large a key as the message which is difficult in practice. Various relaxations from this condition have been explored, including computational security against adversaries with limited resources \cite{rivest1978method,daemen1999aes} and information-theoretic security, where Eve's access to information is restricted \cite{wyner1975wire,liang2009information,bloch2011physical}. Further deviations, such as individual secrecy, have also been widely explored \cite{bhattad2005weakly,silva2009universal,cohen2018secure}. Recent research combining information-theoretic methods and computational security notions without compromising communication rate has been shown to provide energy-efficient secure communication systems by linear coding across the data of multiple links and encrypting a portion of it \cite{cohen2021network,d2021post,woo2023cermet}.

However, these relaxations on the perfect secrecy condition result in information leakage between the ciphertext and plaintext. The leaked information from a cryptosystem can be used to evaluate its strength, particularly against side-channel attacks that exploit the physical properties of the implementation, such as power consumption and time. Although extensive research has been conducted on MI analysis on side channels to evaluate the security of cryptographic systems \cite{gierlichs2008mutual,prouff2009theoretical}, application of MI between the plaintext and ciphertext directly in the context of chosen plaintext attack \cite{diffie1977special,merkle1981security} has not been explored as much. Inferring the MI between the plaintext and its ciphertext could potentially reduce the complexity of finding the key, and thus the security level of the cryptosystem \cite{gierlichs2008mutual}.

Many security schemes, especially those providing individual secrecy, also rely on randomness and expect the input messages to be both independent and uniformly distributed. This ensures that patterns in plaintext do not aid eavesdroppers by helping them deduce information from the ciphertext. Encryption schemes like AES Electronic Cipher Block (ECB) mode \cite{dworkin2016recommendation} and those providing individual secrecy can be susceptible to vulnerabilities arising from such patterns, potentially leaking additional information that assists adversaries in learning more about the combination of input messages. Furthermore, using the same encryption key for multiple encryption instances, especially with large files, can lead to an increase in the mutual information between the plaintext and ciphertext. From an information theory perspective, the only missing element is the key itself. Nevertheless, when dealing with properly uniform data and employing secure encryption methods, the ciphertext may appear entirely uncorrelated with the plaintext. Thus, the MI analysis between plaintext and ciphertext is of special interest in such cases but has been proven challenging due to their high-dimensional and nonlinear relationship. Recent advances in machine learning offer practical ways to estimate MI, using stochastic gradient descent over neural networks \cite{belghazi2018mutual}. This estimation is useful for evaluating the strength of the cryptosystem since it can be exploited by malicious users.

In this article, we explore the use of MI estimation using neural networks to identify weak cryptosystems and their potential use in chosen plaintext attacks. Furthermore, we discuss the different assumptions from the information-theoretic aspect of the cryptosystem and analyze the impact of input uniformity on security. We also use the novel MI estimator using neural networks (MINE) \cite{belghazi2018mutual} to check information leakage due to the partial encryption scheme presented in the Hybrid Universal Network Coding Cryptosystem (HUNCC) \cite{cohen2021network,cohen2022partial} and present the necessary guidelines for its practical use. We test the efficiency of such estimators in identifying the correlation between the plaintext and ciphertext without access to the key 
and its application as a tool for evaluating the security of cryptographic systems.

\section{Neural Estimation of Mutual Information on Cryptosystems} \label{CMINE}

MI between variables with unknown distributions has been prohibitively difficult to calculate, making it an even bigger challenge to calculate MI for a finite dataset of high dimensional plaintext ciphertext pairs. However, the neural estimation of MI proves to be converging to a lower bound as the number of samples goes to infinity \cite[Theorem 2]{belghazi2018mutual}. This process follows the notion that MI between two random variables $X$ and $Y$ can be expressed by the Kullback-Leibler (KL) Divergence, the distance between their joint distribution and the product of their marginal distributions.
\begin{equation*}
I(X;Y) = D_{KL}(P(X,Y) || P(X)P(Y))
\end{equation*}
Donsker-Varadhan representation of KL divergence is used to estimate MI, where $\Omega$ is the product sample space of the distributions $P_1$ and $P_2$, and the supremum is taken over all functions $F$, with a finite expectation. As seen in \cite{belghazi2018mutual}, $F$ can be modeled as a neural network $F_{\phi}$, where $\phi$ is optimized such that a maximum of $I_{\phi}(X;Y)$ can be computed using stochastic gradient descent (SGD) \cite{goodfellow2016deep}.
\begin{equation*}
D_{KL}(P_1 || P_2) = {\sup_{F: \Omega \to \mathbb{R}}} E_{P_1}[F] - \log( E_{P_2}[e^F] )
\end{equation*}
\begin{equation*}
I_{\phi}(X;Y) = E_{P(X,Y)}[F_{\phi}] - \log( E_{P(X)P(Y)}[e^{F_{\phi}}])
\end{equation*}
However, challenges have arisen from high variance with large MI estimations when solving the equation above. To mitigate this problem, \cite{choi2020regularized} added a stabilization term to fix this problem. 

\vspace{-0.3cm}
\begin{multline*}
I_{\phi}(X;Y) = E_{P(X,Y)}[F_{\phi}] - \log(  E_{P(X)P(Y)}[e^{F_{\phi}}]) \\ 
 - 0.1 (\log(E_{P(X)P(Y)}[e^{F_{\phi}}]))^2
\end{multline*}

This regularization term helps the optimizer find one solution for our estimation, rather than wandering between a class of several functions. A more detailed discussion on the MI estimation can be found in \cite{belghazi2018mutual,choi2020regularized}.


\begin{algorithm}
\caption{MI Estimation for Cryptosystems}\label{alg:cap}
\begin{algorithmic}
\State 1: \textbf{Input} Plainext \textbf{X} for \textbf{N} samples
\State 2: \textbf{Encrypt(x)} for ciphertext \textbf{Y} for \textbf{N} samples
\State 2: Initialize network parameters \textbf{$\theta$}
\State 4: \textbf{repeat}
\State 5: \hspace{0.25 cm} Find \textbf{I(X;Y)} between the sample set
\State 6: \hspace{0.25 cm} Compute SGD optimizing and updating \textbf{$\theta$}
\State 7: \textbf{until} convergence
\end{algorithmic}
\end{algorithm}
\vspace{-0.2 cm}

\subsection{Neural Estimation Procedure}
Mutual information neural estimation has been successfully used in different applications and optimizations of MI \cite{belghazi2018mutual,esfahanizadeh2023infoshape,atashin2021variational} and we demonstrate the estimator's capabilities with cryptographic protocols. To set the baseline, a few extreme cases are considered in section \ref{baseline_section} and the experiments are further extended to several popular cryptographic protocols in section \ref{crypto_MI}.

The estimation of MI in our experiments is achieved using the gradient descent of a neural network that takes the batches of the plaintext-ciphertext pairs as the input \cite{belghazi2018mutual,choi2020regularized}, as shown in alg.~\ref{alg:cap}. Specifically, the network consists of two intermediate layers of 100 nodes and an output node providing the MI. The number of input nodes depends on the dimensions of the input to the neural network. Our baseline experiments require 32 nodes in the input layer. We use ReLU non-linearity between layers. For each estimation, we use 100,000 samples. The plaintext and ciphertext each have a length of 16 bytes in our initial set of experiments. We use batch size 10,000, learning rate 1e-4, and 2000 or 5000 epochs, depending on the complexity of the protocol.

\subsection{Baseline Experiments}\label{baseline_section}
For our baseline experiments to analyze the efficiency of the neural estimation, we deploy it on a few baseline scenarios such as plaintext and ciphertext being exactly the same (No encryption), one-time pad encryption, one-time pad with key, and simple XOR with the same key. 
We can get the true MI of no encryption setting, by using $I(X;Y) = H(X) - H(X|Y)$, where $H$ is the Shannon entropy. Consider $X$ to be our plaintext and $Y$ to be our ciphertext. Since we do not encrypt, $X = Y$, and by definition $H(X|Y) = 0$ when $Y = X$. Since our inputs are uniformly generated, we can calculate $I(X;Y) = H(X) \simeq 11.09$~nats\footnote{Our estimator uses natural logarithm} and use this as an upper bound and performance indicator. Similarly, for a completely uncorrelated plaintext and ciphertext scenario, $H(X|Y) = H(X)$ and $I(X;Y) = 0$ by principle. However, the estimator is able to optimize the MI over the finite dataset and estimates $I_{\phi}(X;Y)$ a value negligibly greater than zero.

\begin{figure}[!t]
    \centering
    \includegraphics[width=0.95\linewidth]{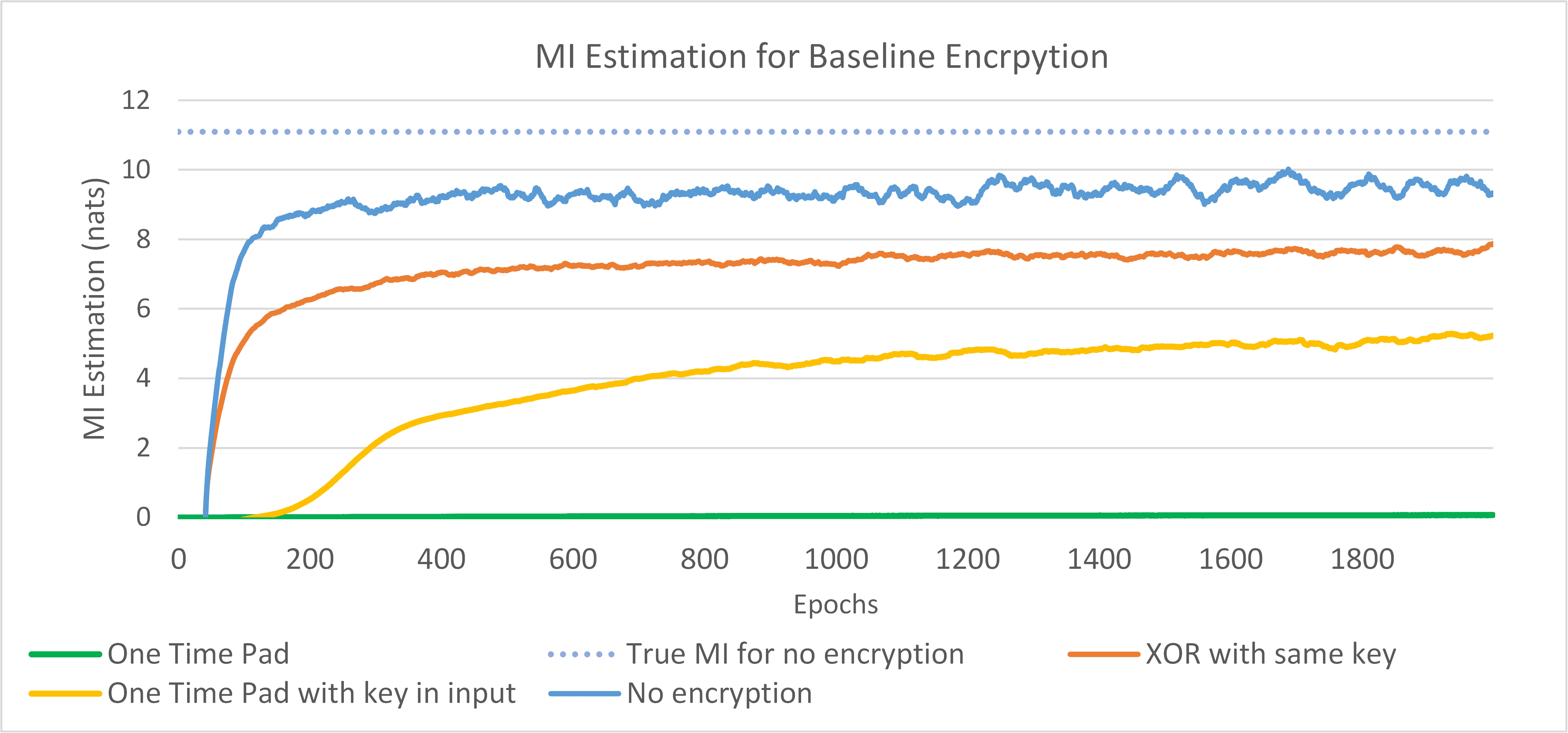}
    \caption{Baseline MI estimation results}
    \label{fig:MI_Baseline}
    \vspace{-0.1cm}
\end{figure}

Our experiments with neural estimation for no encryption setting provide the MI estimation of $I_{NO}(X;Y) = 9.7$ nats converging around 100 epochs, as shown in Fig.~\ref{fig:MI_Baseline}. This shows that the MI estimation of our 16 dimension variables performs with great accuracy. The MI estimation of the repeating key XOR results in $I_{XOR}(X;Y) = 7.8$ nats. This high MI estimation matches our expectations since XOR-ing with the same key is a simple operation that results in a high correlation between the two variables. For our baseline estimation of a strong encryption scheme, we use a one-time pad (OTP), proven to provide perfect secrecy, i.e., zero MI between plaintext and ciphertext~\cite{shannon1949communication}. The MI estimation for the OTP is approximately $0.05$ nats. However, if we add the key used in OTP to the input, the estimator is easily able to detect the correlation of $X$ and $Y$, as shown in Fig.\ref{fig:MI_Baseline}. 

\subsection{MI estimation for popular cryptographic protocols} \label{crypto_MI}

We now use our estimator on several some well-known encryption schemes, including AES ECB, AES Counter Mode (CTR), a single SPN block cipher, and Caesar cipher (stream cipher). We again use the same parameters for the neural network, but due to the complexity of the operations in these encryption schemes, we run the estimator for 5000 epochs. 
 
\begin{figure}[!t]
    \centering
    \includegraphics[width=0.95\linewidth]{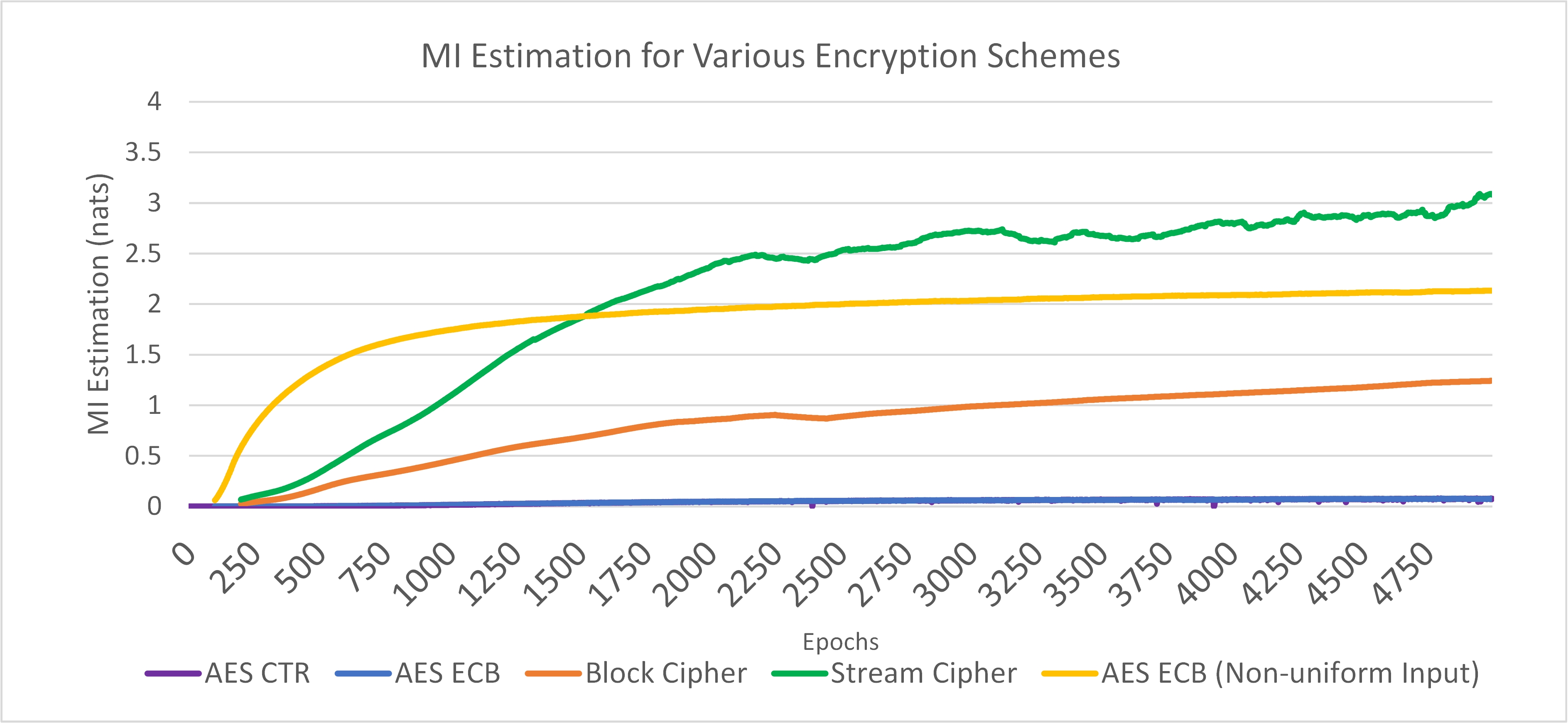}
    \caption{MI estimation for popular cryptographic protocols}
    \label{fig:MI2}
    \vspace{-0.4cm}
\end{figure}

Fig.~\ref{fig:MI2} depicts the results. As can be seen, the block cipher encryption scheme converges to an MI estimate of $I(X;Y) =1.1$ nats, while the stream cipher encryption scheme converges to an estimate of around $I(X;Y) =3.1$ nats. However, estimation over AES ECB and CTR modes result in a very low value, $I(X;Y) = 0.07$ nats. Interestingly, if we try AES ECB with non-uniform, correlated inputs, we estimate an MI of $I(X;Y) = 2.1$ nats. MI leakage from AES ECB implies that the estimator is able to rightly identify leakage of correlation between the plaintext and ciphertext, even though for any particular instance the encryption is secure. 



\section{Mutual Information Analysis of Network Coding Based Cryptosystems}

Cryptosystems that guarantee individual secrecy through coding schemes combined with information theoretic approaches have been of particular interest in achieving secure communication with high data rates. Such systems may leak information about combinations of inputs while protecting each individual message from being decrypted. In this section, we analyze the security of a recently proposed Hybrid Universal Network Coding Cryptosystem (HUNCC) that provides \emph{individual computational security} through coding and partial encryption~\cite{cohen2021network}. This approach is best explained in a network setting with $n$ messages to be sent over $n$ communication links. The messages are linearly encoded using a generator matrix $\textbf{G}\in F^{n \times n}_{2^n}$ using a network coding scheme \cite{silva2009universal,cohen2018secure} before encrypting a subset of the outgoing links with any particular cryptosystem. By encrypting a small portion of the outgoing message, this approach achieves individual and computational security as presented in \cite[Theorem 1]{cohen2021network} and individual indistinguishability under chosen ciphertext attack (IND-CCA1) as in \cite{cohen2022partial}. 

The MI analysis of this scheme is particularly interesting since it uses well-known cryptosystems to achieve computational security of individual messages with the random linear encoding of inputs. Our neural network estimator models a computationally limited adversary that estimates the MI. Furthermore, there are a few similar works that follow the same principle of linear encoding, but instead of cryptography using physical layer security, to achieve absolute physical layer security in high-frequency communication systems~\cite{cohen2023absolute}. The results from this analysis can also be extended to such works with minor modifications.

In this section, we empirically examine these claims and examine HUNCC under different input distributions to analyze the necessary uniformity in HUNCC's input distribution. For this section, the encoding operation for HUNCC follows random linear coding \cite{silva2009universal} and the encryption scheme is AES-128 ECB, though HUNCC's universality allows almost any encryption scheme after encoding.

\subsection{Analysis on HUNCC for Individual Secrecy}

For the initial analysis of the HUNCC scheme, we consider 100,000 samples of uniformly distributed plaintexts of 128 bits, defined in $GF(2^8)$ as 16 bytes. There are 8 outgoing links with only one link encrypted using AES-128 ECB mode. The MI estimate of this setting is indistinguishable from a completely AES-ECB encrypted case, as shown in Fig.~\ref{fig:HUNCC_AES_COMPARISON}, for a uniformly distributed input. We also analyze the individual indistinguishability of the HUNCC scheme as defined in \cite{cohen2022partial}. For this analysis, one particular input message among the 8 input plaintexts is set to a non-uniform input (as 128 bits of 1 in our case) and the rest of the inputs are chosen as random. It was observed that the ciphertexts of each instance were entirely different, and the MI estimator was not able to distinguish any particular pattern, estimating an MI of $I(X;Y) = 0.052$ nats, similar to the case when inputs are uniformly distributed in fig.~\ref{fig:HUNCC_AES_COMPARISON}. This analysis supports the claims in \cite{cohen2021network,cohen2022partial} about individual computational security and individual IND-CCA1 attacks.


\subsection{HUNCC's Robustness to Input Distribution}
We analyze the security of HUNCC under inputs lacking uniformity and compare it with its underlying cryptosystem. We verify this by testing our MI estimator on HUNCC under different levels of input uniformity, ranging from completely non-uniform all the way to completely uniform. We also test AES ECB and AES CTR full encryption for the same inputs to provide a comparison to the state-of-the-art encryption schemes.

\begin{figure}
    \centering
    \includegraphics[width=1\linewidth]{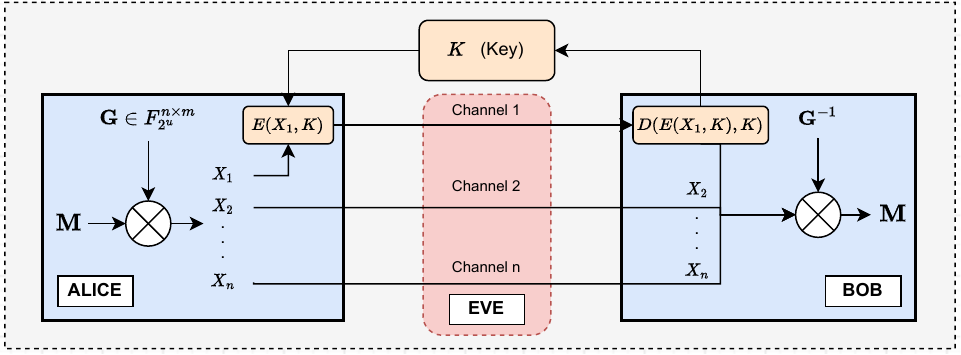}
    \vspace{-0.5cm}
    \caption{Diagram of HUNCC}
    \label{fig:HUNCC}
    \vspace{-0.2cm}
\end{figure}


To create inputs with such varying uniformity, we use a Gilbert-Elliot (GE) model \cite{gilbertpaper,elliottpaper}, with a $1$ and $0$ state, each generating their respective bit. This two-state Markov chain model with a state change probability of alpha ($\alpha$) in both directions for simplicity. We vary $\alpha$ from 0.01 (non-uniform distribution) to 0.5 (uniform distribution) to introduce randomness in the input. For the experiments in this section, consider eight channels, each with 128 bits. Once these inputs are generated by our GE model, they are concatenated into 16 bytes per channel, just as the values were for the encryption schemes in section \ref{CMINE}. This leaves 128 symbols for the input (16 bytes over 8 channels). These inputs are then passed through our \textbf{$\textbf{G}$} encoding matrix, and then the first channel of the eight is encrypted by AES. For comparison, the uncoded inputs are fully encrypted with both AES ECB and AES CTR modes across all eight channels. For each estimation, we use the same parameters and architecture for our neural network as we did throughout section \ref{CMINE}, besides including 500,000 samples and 256 input nodes for the larger inputs.

\begin{figure}
    \centering    
    \includegraphics[width=0.95\linewidth]{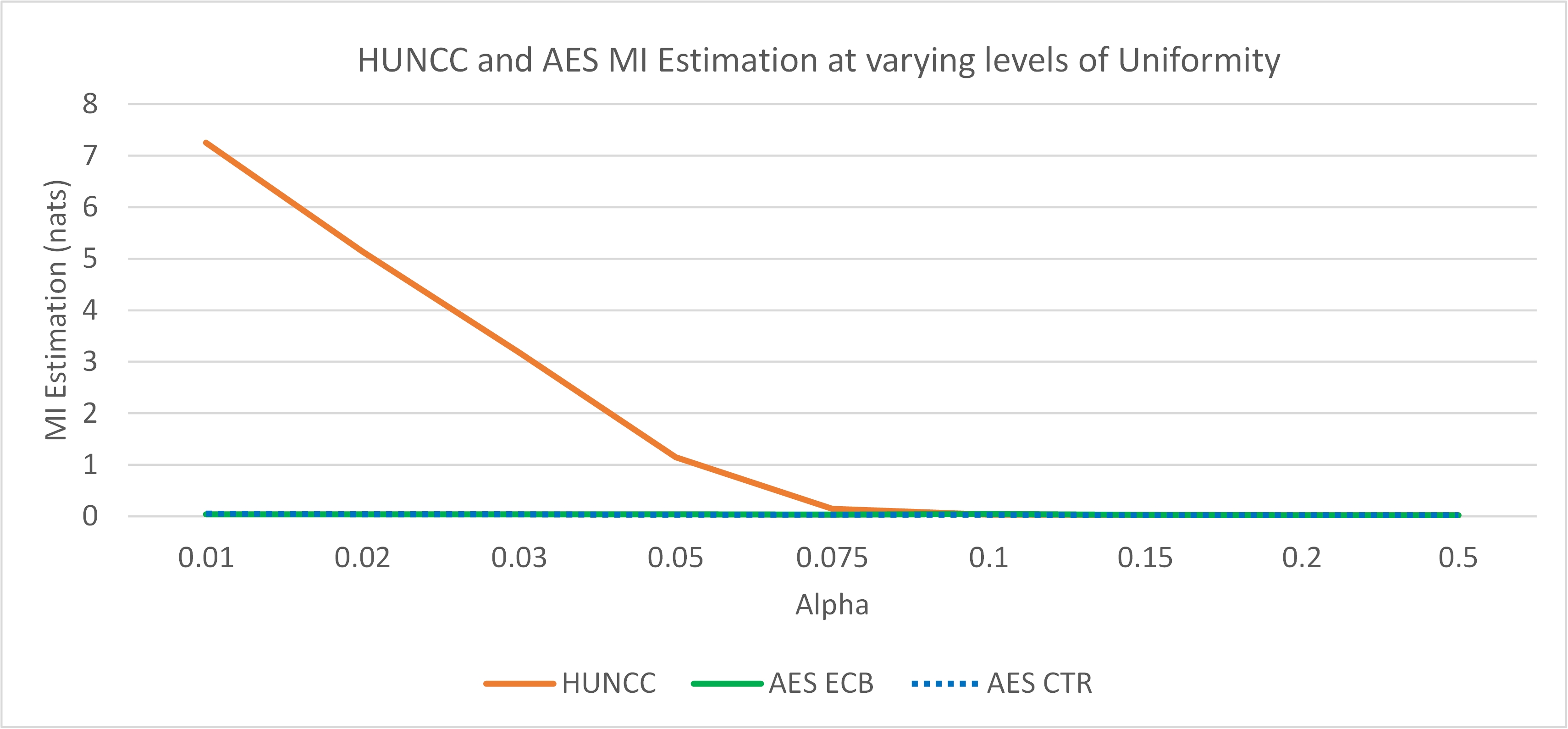}
    \caption{Comparison between HUNCC and AES MI estimation for different levels of uniformity}
    \label{fig:HUNCC_AES_COMPARISON}
    \vspace{-0.3cm}
\end{figure}

Fig. \ref{fig:HUNCC_AES_COMPARISON} and table \ref{table:mi} illustrate the results of the complete set of tests. It is evident that the HUNCC system exhibits significant information leakage when the inputs are non-uniform. However, the MI between inputs and the HUNCC output decreases rapidly as the randomness of the input increases. In fact, HUNCC performance matches that of AES with only a small amount of randomness in the input, an alpha value of 0.1. This analysis suggests that the strict theoretical requirement of a uniformly distributed input may be relaxed when going up against attacks based on learning MI between the plaintext and ciphertext. From a practical point of view, this relaxed requirement can be satisfied by a lossless compression scheme, such as the Lempel–Ziv–Welch (LZW) compression \cite{welch1984technique}. Analyzing the entropy in the inputs to determine randomness, an LZW compressed input with $\alpha$ = 0.02  provides an average Shannon entropy of 1.33 nats while an $\alpha$ = 0.1 input (where MI estimation of HUNCC approximately matches AES), measures an average entropy of 1.52 nats. Since compression schemes are used widely in communication systems for reduced bandwidth usage, this can be achieved without incurring any additional cost. 

\vspace{-0.2cm}
\begin{table}[!t]\small
    \centering
    \caption{\small MI Estimation for Different Levels of Uniformity (nats).}
    \label{table:mi}
    \vspace{0.2cm}
    \begin{tabular}{|l||l|l|l|}
    \hline
        \textbf{$\alpha$} & \textbf{HUNCC} & \textbf{AES CTR} & \textbf{AES ECB}  \\ \hline \hline
        0.01 & 7.2542 & 0.0384 & 0.0536  \\ \hline
        0.02 & 5.1399 & 0.0411 & 0.0411  \\ \hline
        0.03 & 3.1801 & 0.0395 & 0.0384  \\ \hline
        0.05 & 1.1455 & 0.0374 & 0.0263  \\ \hline
        0.075 & 0.1447 & 0.0354 & 0.0281  \\ \hline
        0.10 & 0.0319 & 0.0452 & 0.0298  \\ \hline
        0.15 & 0.0302 & 0.0258 & 0.0371\\ \hline
        0.20 & 0.0233 & 0.0377 & 0.0334  \\ \hline
        0.5 & 0.0270 & 0.0359 & 0.0324 \\ \hline
    \end{tabular}
    
\end{table}



\section{Conclusions}
An accurate MI estimator capable of measuring the leakage between high-dimensional random variables, such as plaintext and ciphertext, can be advantageous in gauging the efficiency of a cryptosystem. Leaked information varies significantly depending on the cryptosystem and the input distribution. Our empirical analysis with MINE showcases the capability of the neural network-based estimator to identify leakage in weaker cryptosystems and its limitations in learning about stronger systems such as AES. This could be an important tool in the cryptanalysis of a security protocol to model a computationally limited adversary, e.g., in chosen plaintext attacks. Furthermore, we investigate how the input distribution impacts the security of cryptosystems, particularly those providing individual secrecy through a combination of linear coding and encryption, as in HUNCC. It is evident that highly correlated inputs result in leakage of information in such systems, but with proper uniformization of the input, the network coding-based cryptosystems limit their MI leakage to the same level as their underlying cryptosystems. Furthermore, our analysis shows that, for a practical application, lossless compression of the input will be sufficient to provide adequately uniform inputs.

\bibliographystyle{IEEEtran}
\bibliography{refs}

\begin{thebibliography}{10}
\providecommand{\url}[1]{#1}
\csname url@samestyle\endcsname
\providecommand{\newblock}{\relax}
\providecommand{\bibinfo}[2]{#2}
\providecommand{\BIBentrySTDinterwordspacing}{\spaceskip=0pt\relax}
\providecommand{\BIBentryALTinterwordstretchfactor}{4}
\providecommand{\BIBentryALTinterwordspacing}{\spaceskip=\fontdimen2\font plus
\BIBentryALTinterwordstretchfactor\fontdimen3\font minus \fontdimen4\font\relax}
\providecommand{\BIBforeignlanguage}[2]{{%
\expandafter\ifx\csname l@#1\endcsname\relax
\typeout{** WARNING: IEEEtran.bst: No hyphenation pattern has been}%
\typeout{** loaded for the language `#1'. Using the pattern for}%
\typeout{** the default language instead.}%
\else
\language=\csname l@#1\endcsname
\fi
#2}}
\providecommand{\BIBdecl}{\relax}
\BIBdecl

\bibitem{shannon1949communication}
C.~E. Shannon, ``Communication theory of secrecy systems,'' \emph{The Bell system technical journal}, vol.~28, no.~4, pp. 656--715, 1949.

\bibitem{rivest1978method}
R.~L. Rivest, A.~Shamir, and L.~Adleman, ``A method for obtaining digital signatures and public-key cryptosystems,'' \emph{Communications of the ACM}, vol.~21, no.~2, pp. 120--126, 1978.

\bibitem{daemen1999aes}
J.~Daemen and V.~Rijmen, ``{AES} proposal: Rijndael,'' 1999.

\bibitem{wyner1975wire}
A.~D. Wyner, ``The wire-tap channel,'' \emph{Bell system technical journal}, vol.~54, no.~8, pp. 1355--1387, 1975.

\bibitem{liang2009information}
Y.~Liang, H.~V. Poor, and S.~Shamai, \emph{Information theoretic security}.\hskip 1em plus 0.5em minus 0.4em\relax Now Publishers Inc, 2009.

\bibitem{bloch2011physical}
M.~Bloch and J.~Barros, \emph{Physical-layer security}.\hskip 1em plus 0.5em minus 0.4em\relax Cambridge University Press, 2011.

\bibitem{bhattad2005weakly}
K.~Bhattad and K.~R. Narayanan, ``Weakly secure network coding,'' \emph{NetCod, Apr}, vol. 104, 2005.

\bibitem{silva2009universal}
D.~Silva and F.~R. Kschischang, ``Universal weakly secure network coding,'' in \emph{2009 IEEE Inf. Theory Works. on Net. and Inf. Theory}.\hskip 1em plus 0.5em minus 0.4em\relax IEEE, 2009, pp. 281--285.

\bibitem{cohen2018secure}
A.~Cohen, A.~Cohen, M.~Medard, and O.~Gurewitz, ``Secure multi-source multicast,'' \emph{IEEE Transactions on Communications}, vol.~67, no.~1, pp. 708--723, 2018.

\bibitem{cohen2021network}
A.~Cohen, R.~G.~L. D’Oliveira, S.~Salamatian, and M.~M{\'e}dard, ``Network coding-based post-quantum cryptography,'' \emph{IEEE journal on selected areas in information theory}, vol.~2, no.~1, pp. 49--64, 2021.

\bibitem{d2021post}
R.~G.~L. D'Oliveira, A.~Cohen, J.~Robinson, T.~Stahlbuhk, and M.~M{\'e}dard, ``Post-quantum security for ultra-reliable low-latency heterogeneous networks,'' in \emph{MILCOM 2021-2021 IEEE Military Communications Conference (MILCOM)}.\hskip 1em plus 0.5em minus 0.4em\relax IEEE, 2021, pp. 933--938.

\bibitem{woo2023cermet}
J.~Woo, V.~A. Vasudevan, B.~Kim, A.~Cohen, R.~G.~L. D'Oliveira, T.~Stahlbuhk, and M.~M{\'e}dard, ``{CERMET: Coding for Energy Reduction with Multiple Encryption Techniques-- It's easy being green},'' \emph{arXiv preprint arXiv:2308.05063}, 2023.

\bibitem{gierlichs2008mutual}
B.~Gierlichs, L.~Batina, P.~Tuyls, and B.~Preneel, ``Mutual information analysis: A generic side-channel distinguisher,'' in \emph{Int. Works. on Crypto. Hardware and Embedded Systems}.\hskip 1em plus 0.5em minus 0.4em\relax Springer, 2008, pp. 426--442.

\bibitem{prouff2009theoretical}
E.~Prouff and M.~Rivain, ``Theoretical and practical aspects of mutual information based side channel analysis,'' in \emph{Applied Cryptography and Network Security: 7th International Conference, ACNS 2009, June 2-5, 2009. Proceedings 7}.\hskip 1em plus 0.5em minus 0.4em\relax Springer, 2009, pp. 499--518.

\bibitem{diffie1977special}
W.~Diffie and M.~E. Hellman, ``Special feature exhaustive cryptanalysis of the nbs data encryption standard,'' \emph{Computer}, vol.~10, no.~6, pp. 74--84, 1977.

\bibitem{merkle1981security}
R.~C. Merkle and M.~E. Hellman, ``On the security of multiple encryption,'' \emph{Communications of the ACM}, vol.~24, no.~7, pp. 465--467, 1981.

\bibitem{dworkin2016recommendation}
M.~Dworkin \emph{et~al.}, ``Recommendation for block cipher modes of operation: methods for format-preserving encryption,'' \emph{NIST Special Pub.}, vol. 800, p. 38G, 2016.

\bibitem{belghazi2018mutual}
M.~I. Belghazi, A.~Baratin, S.~Rajeshwar, S.~Ozair, Y.~Bengio, A.~Courville, and D.~Hjelm, ``Mutual information neural estimation,'' in \emph{International conference on machine learning}.\hskip 1em plus 0.5em minus 0.4em\relax PMLR, 2018, pp. 531--540.

\bibitem{cohen2022partial}
A.~Cohen, R.~G.~L. D’Oliveira, K.~R. Duffy, and M.~M{\'e}dard, ``Partial encryption after encoding for security and reliability in data systems,'' in \emph{2022 IEEE International Symposium on Information Theory (ISIT)}.\hskip 1em plus 0.5em minus 0.4em\relax IEEE, 2022, pp. 1779--1784.

\bibitem{goodfellow2016deep}
I.~Goodfellow, Y.~Bengio, and A.~Courville, \emph{Deep learning}.\hskip 1em plus 0.5em minus 0.4em\relax MIT press, 2016.

\bibitem{choi2020regularized}
K.~Choi and S.~Lee, ``Regularized mutual information neural estimation,'' 2020.

\bibitem{esfahanizadeh2023infoshape}
H.~Esfahanizadeh, W.~Wu, M.~Ghobadi, R.~Barzilay, and M.~M{\'e}dard, ``Infoshape: Task-based neural data shaping via mutual information,'' in \emph{ICASSP 2023-2023 IEEE International Conference on Acoustics, Speech and Signal Processing (ICASSP)}.\hskip 1em plus 0.5em minus 0.4em\relax IEEE, 2023, pp. 1--5.

\bibitem{atashin2021variational}
A.~A. Atashin, B.~Razeghi, D.~G{\"u}nd{\"u}z, and S.~Voloshynovskiy, ``Variational leakage: The role of information complexity in privacy leakage,'' in \emph{Proceedings of the 3rd ACM Workshop on Wireless Security and Machine Learning}, 2021, pp. 91--96.

\bibitem{cohen2023absolute}
A.~Cohen, R.~G.~L. D'Oliveira, C.-Y. Yeh, H.~Guerboukha, R.~Shrestha, Z.~Fang, E.~Knightly, M.~M{\'e}dard, and D.~M. Mittleman, ``Absolute security in terahertz wireless links,'' \emph{IEEE Journal of Selected Topics in Signal Processing}, 2023.

\bibitem{gilbertpaper}
E.~N. Gilbert, ``Capacity of a burst-noise channel,'' \emph{Bell sys. tech. journ.}, vol.~39, no.~5, pp. 1253--1265, 1960.

\bibitem{elliottpaper}
E.~O. Elliott, ``Estimates of error rates for codes on burst-noise channels,'' \emph{The Bell Sys. Tech. Journ.}, vol.~42, no.~5, pp. 1977--1997, 1963.

\bibitem{welch1984technique}
T.~A. Welch, ``A technique for high-performance data compression,'' \emph{Computer}, vol.~17, no.~06, pp. 8--19, 1984.

\end{thebibliography}

\end{document}